# Advanced Caching for Distributing Sensor Data through Programmable Nodes

M. Femminella, G. Reali, D. Valocchi
DIEI – University of Perugia
Via G. Duranti 93, Perugia, Italy
mauro.femminella@diei.unipg.it,
gianluca.reali@diei.unipg.it, dario.valocchi@gmail.com

R. Francescangeli, H. Schulzrinne
Computer Science – Columbia University
New York, USA
roberto.francescangeli@gmail.com, hgs@cs.columbia.edu

*Abstract*—This paper shows an innovative solution for distributing dynamic sensor data by using distributed caches. Our proposal is based on the concepts of service modularization and virtualization of network nodes made available by the NetServ hosting environment, which has been defined and implemented with the aim of extending the functions of the network nodes. Through a lab experiment involving tens of nodes, we have demonstrated a significant performance improvements in term of traffic saving and download time in comparison with a legacy, Internet-based, approach. Beyond this performance improvements, the proposed solution holds also functional improvements, in terms of dynamic deployment and easy integration with services making use of sensor data.

## I. INTRODUCTION

The evolution of the applications using sensor data is driven by the availability of an increasing quantity of information types. Any device collecting information from the surrounding environment can be regarded as a sensor. Hence, sensor data could be both the measure of environmental physical quantities, such as temperature and humidity, and other information obtainable through the peripherals of modern portable devices, such as smart-phones or wearable devices.

This paper shows a proposal for making the information collected from a number of distributed sensors available to a large number of recipients in a close to real time fashion.

Our proposal was stimulated by the diffusion of new applications having the following features:
- Applications make use of highly variable data.
- Data are collected by geographically distributed sensors.
- The number of expected users is large and their positions are spatially correlated.
- Sensor data need to be integrated within a complex service architecture involving different entities.

Examples of such applications are crowd tracking, navigation services, and augmented reality gaming [18].

These services request the transfer of a considerable amount of data from many sources to many destination. For this reason they can pose some challenges to the core of the involved networks and suitable traffic engineering techniques are needed. Hence, the aim of our proposal is to facilitate the retrieval of data collected by sensors and their integration with other software components making use of them (e.g., augmented reality servers, security engines, fault recovery applications and so on).

The system architecture is sketched in Fig. 1, which depicts all involved entities and their mutual logical relations.

We assume that a number of sensors, ranging from few units to several thousands [15], can collect any given type of information and store it within remote nodes, labeled as Gateways. For the sake of clarity, we stress that most of the literature on sensor networks is focused on this data transfer. Differently, our solution is focused on the transfer of sensor data to applications and their usability. Thus, whilst our experimental activity makes use of well assessed solutions for transferring and storing data within Gateways, the original and innovative contribution is within the procedure to deliver sensor data from the Gateway to applications through the network.

The data transfer service to applications is specified in terms of quality of service (QoS), intended as sufficiently small response time to queries, recency, meaning that data values read through a query transaction must be sufficiently close to those relevant to the query time, and consistency, meaning that values read must be either valid at the time of reading or relevant to times sufficiently close each other [9]. Our solution makes use of node virtualization and advanced signaling. It builds on the achievements of the NetServ research project[1], lead by the Columbia University [4]. In more detail, it consists of dynamically deploying caching module within the NetServ-enable nodes lying on the path between the Gateway and the remote client requesting sensor data. In this way, a sort of dynamic, soft-state, content delivery network (CDN) is created. However, the differences with legacy CDNs are remarkable. In fact, even if most CDN providers have cache nodes deployed close to the edge of the network, that is close to requesting clients/applications, these caches have to be populated in advance with contents potentially popular. In the case of rapidly changing sensor data, this would imply to continuously preload a large amount of data into these nodes, without any guarantees that they would be requested by anyone within the time frame in which they are of any interest. Instead, the proposed solution does not preload anything, and routers acting also as caches would be crossed by the data traffic anyway, belonging it to the data path. The only overhead is represented by the signaling traffic necessary to deploy these software modules, which is shown to be negligible.

Although we will show an experiment dealing with a specific application, our proposal do not assume any restriction to the usage of information collected by sensors. We rather aim to deploy a dynamic service architecture that enables instantiating and de-instantiating service modules in network nodes according to service logic and variable conditions, so as to allow users having the maximum benefits from a service through advanced information retrieval mechanisms. The lack of restrictions in designing and deploying services in user terminals has induced us to leave them unaware of the NetServ-based distribution architecture. Therefore, the scope of our proposal does not include user terminals, although their inclusion would have provided further options to achieve our design objectives. In addition, we do not pose any constraints or requirements on sensors, which operate independently on the presence of the proposed delivery architecture.

[1]The NetServ project is funded by the National Science Foundation under grant NSF-CNS #0831912 as a part of its Future Internet Design (FIND) initiative, and also by DOCOMO Communications Laboratories Europe.

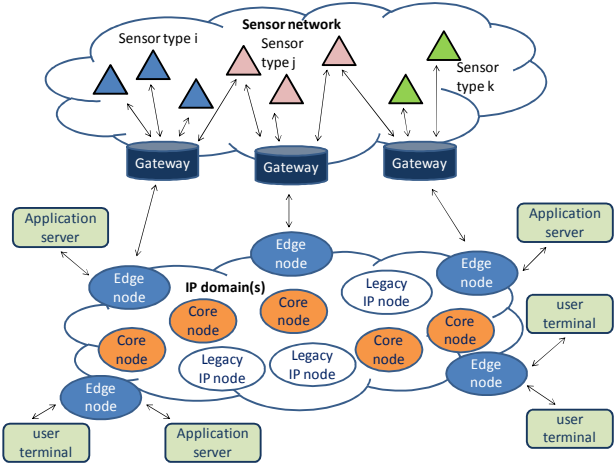

Fig. 1. Network scenario, core and edge nodes are NetServ-enabled nodes.

In order to illustrate all implemented functions within each entity in Fig. 1 it is necessary to describe the state of the art of the involved technologies. This is indeed the contents of Section II. After that, in Section III we will use the concepts introduced to detail all network operations that allow user applications to efficiently retrieve information collected from different sensors, regardless their physical location. The benefits of our solutions have been demonstrated experimentally, by using tens of nodes implementing the entities shown in Fig. 1, along with the protocols supporting their mutual relations, as shown in Section IV. The review of the related work in the field is presented in Section V. Some final comments are reported in Section VI.

## II. BACKGROUND ON TECHNOLOGIES AND PROTOCOLS

### A. NetServ: virtualized network services

NetServ is an NSF research project related to the future Internet. NetServ routers are not limited to implementing routing functions. Through an extensive use of the virtualization concept and dynamic installation and removal of service modules, all network nodes may host components of the service architecture [4].

Fig. 2 shows the software architecture of a network node implementing the NetServ hosting environment. When a signaling message passes through a NetServ router, it may cause it to download and install a module from an available repository. In our experiments, illustrated in Section III, the NetServ repository is implemented in Gateways. The events that trigger signaling and the relevant actions depend on the implemented application. Our NetServ prototype is based on the Linux operating system. Signaling packets are intercepted by the signaling daemons, which extract the signaling content and pass it to the NetServ controller. According to its content, this controller issues commands to the appropriate service containers, such as for installing or removing service modules.

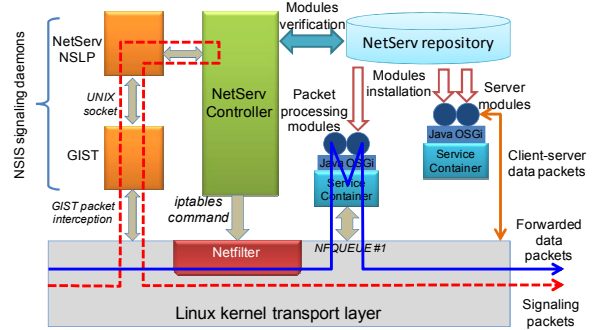

Fig. 2. NetServ node architecture.

Service containers are user space processes. They include Java Virtual Machines (JVMs) which can handle multiple service modules. JVMs execute the OSGi framework [20], hence application modules consists of OSGi-compliant JAR files, referred to as bundles. Since the OSGi framework allows bundles to be loaded and unloaded while the JVM is running, the NetServ application modules can be installed and removed at runtime. Our decision of using Java is due to its code portability, library availability, and support.

As shown in Fig. 2, NetServ considers two types of application modules, the server application modules, and the packet processing application modules. The server application modules behave as standard network servers which communicate with the outer world through the TCP/IP stack (see the orange arrow in Fig. 2). The packet processing modules can process data packets in NetServ nodes composing the data path (see the blue arrow in Fig. 2) [6]. Clearly, the distinction between server and packet processing modules is logical. Each application module could be of both types. Thus, the traditional distinction between a router and a server is no longer necessary. Fig. 2 shows also the adoption of the NSIS signaling architecture [1], illustrated in some details in the next sub-section due to its important role in our proposal.

### B. The NSIS signaling

The NSIS signaling protocol suite, shown in Fig. 3, is organized in two layers [1]. The lower layer, the NSIS Transport Layer Protocol (NTLP), has a transport role. It has to deliver higher-layer signaling messages from a signaling node to the next one on the path to the destination. The General Internet Signaling Transport (GIST) protocol [3] is the most common NTLP implementation. It allows discovering the next-hop NSIS node, which might not correspond to the next routing hop in case some legacy nodes, NSIS ignorant, are included in the path to the destination. The signaling logic serving the application is implemented in the highest layer of the NSIS stack, referred to as NSIS Signaling Layer Protocol (NSLP). Since NSIS nodes are incrementally deployed in real networks, they tend to be sparse and for this reason they can be regarded as an overlay network. GIST does not define new transport protocols or security mechanisms, but rather it makes use of the existing protocols. Applications can specify the desired transport attributes for the signaling flow, so that GIST can select the most appropriate transport protocol(s) to satisfy the specified requirements. Typically, UDP is used when unreliable signaling is acceptable, otherwise TCP is selected, possibly along with TLS for achieving a secure signaling exchange. Extensibility functions are also included in GIST, in order to allow the usage of different transport protocols [2].

The scope of the GIST protocol embraces adjacent nodes only. The end-to-end logic is included in the transported NSLP messages. In this work, we used the NSLP developed in the NetServ architecture, described in some detail in [6].

Routing of signaling messages is governed by the Message Routing Method (MRM), which specifies the algorithm used by GIST to discover the other NSIS compliant nodes and route signaling messages. The standardized GIST specifications define two MRMs:

- the Path Coupled MRM, designed to propagate the signaling messages through the data path;
- the Loose End MRM, used for preconditioning the states in firewalls and NATs along the data path.

In our experiments we have used the Path Coupled MRM. NSIS has been used to trigger installation of storage modules for sensor data within nodes along the data path from the Gateways to the user terminals.

GIST messages deliver also Message Routing Information (MRI) objects that allow NSIS nodes to identify the MRM used and to route signaling messages. For example, in the case of a Path coupled MRM, GIST packets are intercepted by NSIS nodes along the path and then re-injected into the network after being processed by the NSLP entities. Re-injected packets are routed towards the destination, and might be intercepted again by other NSIS-capable nodes before arriving at the desired endpoint, as it happens in our proposal.

The current implementation of the NetServ daemons is based on an extended version of NSIS-ka, an open source NSIS implementation by the Karlsruhe Institute of Technology [5]. In addition to the supported MRMs (see also [14]), GIST can be extended in almost all other internal entities.

III. NETWORK ARCHITECTURE FOR SENSOR DATA DELIVERY

The aim of the sensor network shown in Fig. 1 is to store sensor data in Gateways, so that they can be distributed to applications. A lot of research on sensor networks focus on this problem. Nevertheless, it is not part of our proposal, which is focused on the usability and integration of sensor data with other applications. However, in order to demonstrate the effectiveness of our proposal by using real sensors in operation, for realizing experiments we used data collected by wireless sensors motes Zolertia Z1, equipped with an IPv6 over IEEE 802.15.4 network interface. We used the well-known IETF Constrained Application Protocol (CoAP) [7] to transfer data. These motes execute a CoAP server which sends the collected data to a requesting Gateway, which act as a CoAP client. The CoAP client has been implemented as a NetServ bundle by using the Californium Java implementation [8].

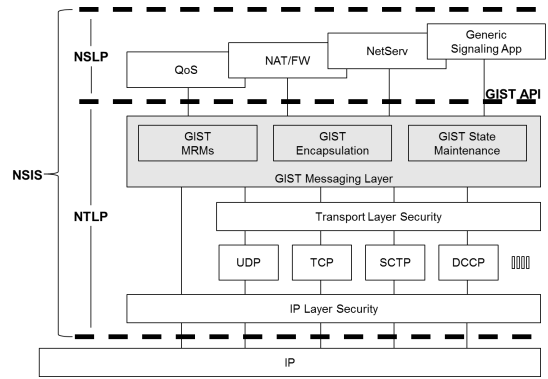

Fig. 3. NSIS framework: 2-layers architecture with detailed transport and supported GIST entities.

In our experiment we have used a single Gateway, since the use of multiple Gateways does not provide any functional advantage. This Gateway has a central role since it provides two sets of functions. Through the interface with the sensor network, it issues data requests to sensor motes, or register with them to receive updates only upon the occurrence of specific events [7], receives data, and stores them within an internal data base (DB). Through the Internet interface, the Gateway allows applications to request the stored data. As for the data distribution, we have implemented, inside the same bundle, a restful interface by using Jetty [16], and formatted the sensor data payload according to the JSON [17] model. This interface is also used to exchange messages with the other NetServ-compliant nodes. In particular, it can initiate NSIS signaling sessions for installing application modules within all NetServ nodes along a specific path.

As for the NetServ application used for distributing sensor data, we have adapted a Content Distribution Network (CDN)-like application that has already provided good results for distributing other data types [4]. In general, CDN solutions provide significant benefits in large networks by replicating data in mirrors close to the expected users. Nevertheless, this solution is not efficient in case of rapidly changing data and a dynamic request patterns, as it happens in the class of applications considered in this paper. Our solution has the advantage of dynamically deploying cache modules on request by applications, without the overhead of preloading contents. We have realized a NetServ bundle implementing CDN-like functions, referred to as Sensor_CDN. Differently from traditional CDNs, all functions are dynamical, data are not bound to specific locations, and sensor data can be replicated also in nodes very close to user terminals. The Sensor_CDN original bundle is stored in the Gateway (see the NetServ repository depicted in Fig. 2). Its internal logic allows both storing data coming from an upstream cache into the DB and sending the data towards the downstream cache/client.

In addition to storing the JSON payload as a single file, the bundle unpacks the data organized in the JSON payload and stores the entries relevant to each single sensor in the DB. This is another improvement compared to the traditional CDNs, since it gives the possibility of fulfilling the data requests by combining the still valid sensor data already stored in caches with some missing data downloadable from the Gateway. This ability improves scalability and allows saving a considerable amount of network resources. In fact, clients organize queries by using a set of filtering rules in the request url so as to specify the type of data they are interested to, the source(s), the time frame, and so on. Thus, if a NetServ node running the Sensor_CDN bundle stores only a portion of the requested data, the mechanism illustrated below allows retrieving only the missing part from the upstream node.

Fig. 4 shows an example of how Sensor_CDN works. When a client requests sensor data from the Gateway, this initiates a NSIS signaling towards the client by using the NetServ NSLP. Only the NetServ nodes on the path are able to understand the message, and this signaling will install the Sensor_CDN bundle on these nodes with the SETUP message [6]. In addition, these nodes communicate both their capabilities and the ordered list of the NetServ nodes along the path to the destination, through a PROBE message. Finally, the Gateway configures the caches in these nodes by another SETUP message, informing the NetServ router 1 to download the content from the NetServ router 2, and the latter to download the content from the Gateway (Fig. 4). Finally, the Gateway issues an HTTP redirect message to the client. Then, the client sends an HTTP GET message to the NetServ router 1, which triggers the download chain towards the Gateway, that will fill the on-path caches while serving the client. If later another client requests the same content, it is redirected to the NetServ router 1, which is now able to send directly the requested data. The Gateway can explicitly send a REMOVE message to uninstall the Sensor_CDN module, otherwise the module will be removed automatically after the number of seconds specified in the TTL field of the SETUP message.

## IV. PERFORMANCE EVALUATION

The achievements of our proposal have been evaluated by using some metrics which highlight its peculiarities. In particular, we have recorded the total network traffic generated by the distribution of a given amount of sensor data. In addition, we have evaluated the average latency for obtaining the requested data. Finally, we have recorded the amount of signaling traffic generated by GIST. Essentially, we have evaluated both the benefits provided by our solutions in terms of the aforementioned QoS, and the relevant costs.

### A. Network testbed

Our experimental testbed is composed of 60 nodes, which are either physical nodes or virtual ones, implemented by virtual machines (VMs). We consider the use of virtual nodes necessary due to both the increasing diffusion of virtualized network solutions and the increased technical challenges that they pose, being them less performing than the host physical nodes. The network topology is composed of 11 core nodes, realized by physical nodes equipped with at least 512 MB of RAM and 1 or 2 CPUs, 12 edge nodes, realized by VMs with 512 MB and 1 virtual CPU, and 37 end-nodes connected to the edge nodes, realized with VMs with 374 MB of RAM and 1 virtual CPU. One end node is used as control node, whereas 35 end-nodes implement the clients requesting the sensor data. Finally, one end node implements the Gateway which is in charge of retrieving data from sensors and distributing them. The network connection of the sensor Gateway is limited to 10 Mb/s, in order to emulate a 3G+/4G cellular connection, which is commonly used for sensor gateways. Edge nodes are connected to at least two core nodes, and some of them are interconnected directly to each other. All nodes have 1 physical network interface, connected to a common switch, and the topology is realized by making use of VLANs. The connections among either edge nodes or between edge nodes and end nodes are Gigabit Ethernet, as it typically happens in local area networks, whereas connections among either core nodes or between core nodes and edge nodes are Fast Ethernet. We have evaluated experimentally that the number of sensors, nodes, and terminals is sufficient to highlight the benefits of our solutions, although even thousands of entities may be involved in operation, thus increasing the evaluated benefits.

### B. Numerical results

We have considered a population of 1000 sensors, which is realistic in large settings [15]. Each sensor sends updates to the Gateway four times each minute. They have a JSON payload size equal to 180 bytes, which includes sensor address and/or identification code, timestamp, sensor coordinates, and sensed data. Assuming that applications ask for data updates each 6 minutes (thus obtaining 24 measurements for each sensor), the overall payload size is equal to about 4 MB. By doubling the network size, the resulting payload is clearly about 8 MB.

Fig. 5 shows the measured download delay values for both 4 MB (Fig. 5.a) and 8 MB (Fig. 5.b), for three different solutions. The first is the legacy one, for which each client downloads data from the Gateway. These sample delay values are all close to 5 seconds, since the network bottleneck is due to the 10 Mb/s link connecting the Gateway to the network. The other two solutions are referred to as "NetServ, E" and "NetServ, E+C". The former consists of deploying the NetServ hosting environment only on edge nodes, whereas the latter consists of deploying the NetServ environment also on one core node located in a strategic position, connected both to the sensors Gateway and to 4 edge nodes. This choice is useful to analyze the possibility of introducing NetServ incrementally in the core. This is a key aspect for the introduction of any new technology, since it allows distributing investments over the time while achieving some benefits since the beginning.

As we can see in Fig. 5, download time values for the NetServ-based solutions are grouped around three values. The first is about 6.5 s, with only one sample. This value corresponds to the first download from a single client. During this download, data are stored at least in two Sensor_CDN cache memories, the one of the edge node directly connected to the sensor Gateway, and the one of the edge node connected to the client requesting the sensor data. Subsequent download times are grouped around two values; most of clients can download the content in less than 2 seconds, since their edge node already stores the requested data, thus exploiting a Gigabit Ethernet connection for a rapid download, whereas all the other clients can download the requested data in about 3-4 seconds. In these cases, the content is downloaded from the edge node connected to the Gateway, thus the bandwidth bottleneck is located in the core connection at 100 Mb/s. In addition, also clients served by an edge node connected to another edge node which stores the sensor data will benefit from the 1 Gb/s link and achieve a download time below 2 seconds. The time needed to store data in the cache memory results negligible. A further observation is that the download time of the two NetServ-based solutions is essentially the same. This is due to the fact that the dominant effect determining the download time is the bottleneck in the network path, whereas the number of network nodes does not have a significant impact, since all nodes are deployed within a LAN. Hence, propagation times are very small and nearly constant. Since the core node is equipped by a Fast Ethernet interface, we expect an improvement only in the amount of the overall network traffic, as shown in what follows.

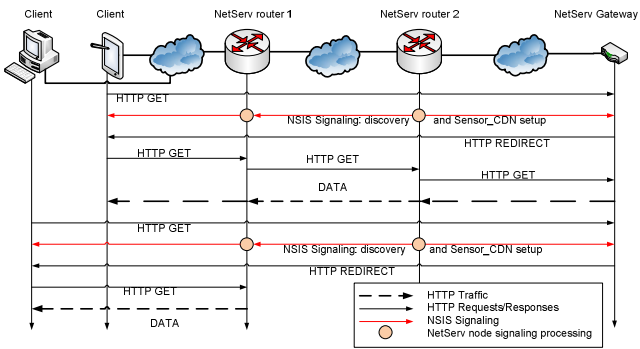

Fig. 4. Signaling exchanges for service execution.

When the aggregate sensor data size is 8 MB, similar considerations apply. Clearly the download times are nearly doubled. This is especially true in the legacy case, whereas in both NetServ compliant solutions the download time is less than twice that obtained by using a package of 4 MB. In order to explain this behavior, it is useful to examine the results shown in Fig. 6, which illustrates the average download time for the three considered solutions and highlights the different contributions to the download time. They are:

- the time taken to create the package, i.e. the time needed to extract data from the DB and create the JSON payload;
- the overhead time associated with the NSIS signaling;
- the processing time, including the HTTP redirection;
- the HTTP data download time.

It is evident that, being the NSIS overhead fixed (about 1.6 seconds), the larger the sensor data payload, the lower the percentage impact on the overall download time. In addition, both processing and HTTP redirect times are negligible (about 10 ms). Finally, the time needed to create the payload is different between the NetServ and the legacy solutions. In fact, in the legacy case, each time a client requests the data, the gateway has to re-create the JSON payload from scratch, and it is time consuming. Instead, in the NetServ cases, since this operation has to be executed once, its impact averaged over 35 requests is negligible (about 16 ms). The gross total indicates that the average time is more than halved for both data sizes.

Fig. 7 shows the amount of traffic handled by the network, being it both the traffic due to the HTTP transfer only, and the total traffic, that is the sum of the HTTP and NSIS contributions. The first observation is that, being the total NSIS traffic equal to 90 KB, its impact is by far negligible in comparison to the HTTP traffic, which is in the order of hundreds of MBs. A remarkable result is quite evident. Whereas the slope of the curve associated with the HTTP legacy case is nearly constant, as indeed expected, the NetServ cases ("NS, E" and "NS, E+C") show two different slopes.

They are relevant to the different behaviors illustrated in the comments to Fig. 5. When the sensor data are downloaded from the caching nodes close to the Gateway, these data have to cross the network, and thus a higher slope results. However, during this phase, the cache close to the client is updated. This will further reduce the total traffic (and thus the slope of the curve) when another client connected to the same edge node will request the content. In both cases, the slope of the NetServ curve is definitely lower than in the legacy case. For 35 requests, a 65% of traffic reduction has been achieved. A final comment is relevant to the introduction of a NetServ-compliant core node. Since this node is connected to 4 edge nodes, during the download from the first client connected to one of these edge nodes, the information stored in its cache is updated. Thus, the time and traffic saving will occur during the subsequent three downloads issued by clients connected to these edge nodes. Further downloads will be directly served by the edge nodes, and the core node is no longer involved. Since the core node used to deploy NetServ is connected to the sensors gateway, the traffic saving amounts to three times the payload size, which is about 12 MB, shown in Fig. 7. Clearly, by deploying NetServ on all core nodes would further increase the network traffic saving without additional costs, since the NSIS signaling crosses core nodes on the data path anyway.

A significant quantity is the amount of computing resources needed to execute the NetServ environment and the bundle handing sensor data. Table I reports the CPU and memory occupation requested by the three main components of NetServ: the controller, the NSIS signaling daemon, and the container hosting the bundles. These values are relevant to three different cases:

- empty NetServ container, signaling daemon running (E);
- peak resource consumption during bundle execution (P);
- Sensor_CDN bundle deployed in the container but idle (I).

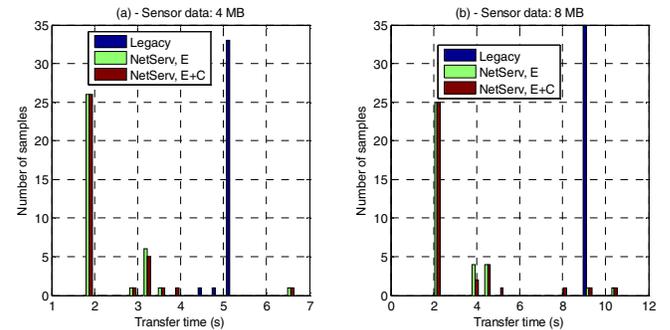
Fig. 5. Download time distribution for different solutions.

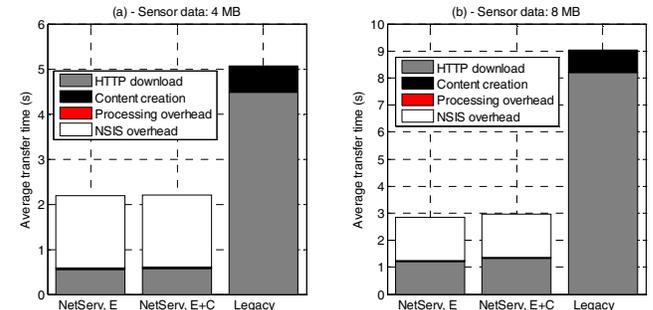
Fig. 6. Average download time with individual delay components.

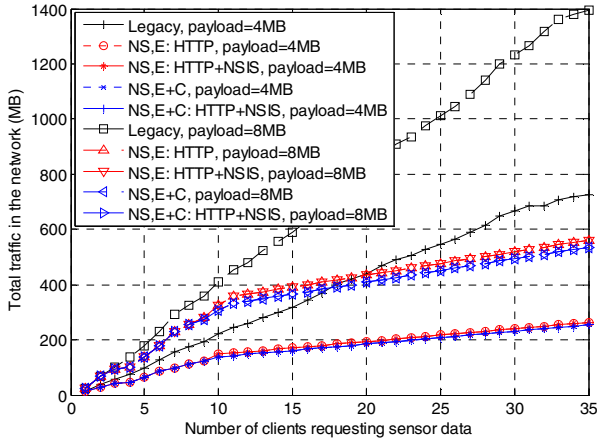

Fig. 7. Network traffic vs. number of clients requesting sensor data for different solutions, size of the JSON payload equal to 4 and 8 MB.

TABLE I. NETSERV RESOURCE CONSUMPTION

| Resource | NSIS | | | Container | | | Controller | | |
|---|---|---|---|---|---|---|---|---|---|
| | E | P | I | E | P | I | E | P | I |
| CPU (%) | 0.7 | 0.7 | 0.7 | 0.2 | 21 | 0.2 | 0 | 0.2 | 0 |
| RAM (MB) | 3 | 4 | 3.5 | 32 | 215 | 117 | 5.6 | 5.6 | 5.6 |

Resources are measured in terms of percentage usage of a single core of an Intel i5 CPU@2.8GHz, and amount of RAM in MB. Data show that the requested resources are definitely affordable by modern router hardware.

## V. RELATED WORK

A lot of proposals and research projects aim to define global scale sensor network that connects sensors to wide-area networks, including the Internet [12][13]. These proposals are essentially focused on functional requirements rather than network performance.

An alternative approach for implementing a real-time spatial data collection system is proposed in [10], where the authors consider data generated by sensors observing real-time situations. The solution proposed is peer-to-peer and each sensor corresponds to a node of an overlay network. This choice is motivated by a new perspective. In fact, the proposed data collection method extends the hierarchical Delaunay overlay network [11] and collects sensor data by avoiding particular sensors according to the required granularity.

A deep analysis of the possible characteristics of sensor storage systems, addressing also scalability problems, can be found in [19], along with a proposal for a storage architecture that envisions data management from metadata by employing local archiving at the sensors and distributed indexing at the proxies. This type of analysis is complimentary to the use of CoAP in the sensor network shown in Fig. 1, the deployment of which can benefit from the insights of [19].

## VI. CONCLUSION

We have shown a dynamic network and application solution for efficiently retrieving data from a large set of sensors. Experimental results show the benefit introduced by using a dynamic programmable nodes paradigm, in terms of download latency and overall network traffic. The benefits of our solution for handling sensor data is not limited to the performance metrics illustrated above. A further important benefit consists of the possibility of handling such data within the same hosting environment where other service components reside. This means that developing services that make use of sensor data does not require any additional integration burden from developers. In addition, sensor-based application can benefit of improved network performance and better QoS.

Future analyses will consider in presence of links with delays typical of geographical networks. In addition, we will improve the caching bundle with packet processing capabilities, in order to intercept data request and limit the NSIS signaling to a very few cases.